\documentclass[conference]{IEEEtran}
\IEEEoverridecommandlockouts
\usepackage{cite}
\usepackage{amsmath,amssymb,amsfonts,hyperref}
\usepackage{algorithmic}
\usepackage{graphicx}
\usepackage{textcomp}
\usepackage{xcolor}
\usepackage[ruled,vlined]{algorithm2e}
\usepackage{booktabs} % For formal tables
\usepackage{subfigure}

\def\BibTeX{{\rm B\kern-.05em{\sc i\kern-.025em b}\kern-.08em
    T\kern-.1667em\lower.7ex\hbox{E}\kern-.125emX}}
\begin{document}

\title{Prototypical Contrastive Learning through Alignment and Uniformity for Recommendation}

\author{
    \IEEEauthorblockN{Yangxun Ou}
    \IEEEauthorblockA{
        \textit{School of Computer Science,}\\
        \textit{East China Normal University,}\\
        Shanghai, China \\
        51215901024@stu.ecnu.edu.cn
    }
    \and
    \IEEEauthorblockN{Lei Chen}
    \IEEEauthorblockA{
        \textit{School of Computer Science,}\\
        \textit{East China Normal University,}\\
        Shanghai, China \\
        lchen@cs.ecnu.edu.cn
    }
    \and
    \IEEEauthorblockN{Fenglin Pan}
    \IEEEauthorblockA{
        \textit{School of Science,}\\
        \textit{Zhejiang University of Technology,}\\
        Hangzhou, China \\
        202105720123@zjut.edu.cn
    }
    \and
    \IEEEauthorblockN{Yupeng Wu}
    \IEEEauthorblockA{
        \textit{School of Computer Science,}\\
        \textit{East China Normal University,}\\
        Shanghai, China \\
        51215901031@stu.ecnu.edu.cn
    }
}

% \author{
%     \IEEEauthorblockN{Anonymous Authors}
%     \IEEEauthorblockA{
%         \textit{dept. name of organization (of Aff.)}\\
%         \textit{name of organization (of Aff.)}\\
%         City, Country \\
%         Anonymous Email Address
%     }
%     \and
%     \IEEEauthorblockN{Anonymous Authors}
%     \IEEEauthorblockA{
%         \textit{dept. name of organization (of Aff.)}\\
%         \textit{name of organization (of Aff.)}\\
%         City, Country \\
%         Anonymous Email Address
%     }
%     \and
%     \IEEEauthorblockN{Anonymous Authors}
%     \IEEEauthorblockA{
%         \textit{dept. name of organization (of Aff.)}\\
%         \textit{name of organization (of Aff.)}\\
%         City, Country \\
%         Anonymous Email Address
%     }
%     \and
%     \IEEEauthorblockN{Anonymous Authors}
%     \IEEEauthorblockA{
%         \textit{dept. name of organization (of Aff.)}\\
%         \textit{name of organization (of Aff.)}\\
%         City, Country \\
%         Anonymous Email Address
%     }
% }

\maketitle

\begin{abstract}
Graph Collaborative Filtering (GCF), one of the most widely adopted recommendation system methods, effectively captures intricate relationships between user and item interactions. 
Graph Contrastive Learning (GCL) based GCF has gained significant attention as it leverages self-supervised techniques to extract valuable signals from real-world scenarios.
However, many methods usually learn the instances of discrimination tasks that involve the construction of contrastive pairs through random sampling. GCL approaches suffer from sampling bias issues, where the negatives might have a semantic structure similar to that of the positives, thus leading to a loss of effective feature representation.
To address these problems, we present the \underline{Proto}typical contrastive learning through \underline{A}lignment and \underline{U}niformity for recommendation, which is called \textbf{ProtoAU}.
Specifically, we first propose prototypes (cluster centroids) as a latent space to ensure consistency across different augmentations from the origin graph, aiming to eliminate the need for random sampling of contrastive pairs. 
Furthermore, the absence of explicit negatives means that directly optimizing the consistency loss between instances and prototypes could easily result in dimensional collapse issues.
Therefore, we propose aligning and maintaining uniformity in the prototypes of users and items as optimization objectives to prevent falling into trivial solutions.
Finally, we conduct extensive experiments on four datasets and evaluate their performance on the task of link prediction. Experimental results demonstrate that the proposed ProtoAU outperforms other representative methods. The source codes of our proposed ProtoAU are available at \url{https://github.com/oceanlvr/ProtoAU}.
% \textcolor{blue}{Finally, we conduct extensive experiments on Yelp and Amazon datasets and evaluate its performance on the task of link prediction. Experimental results demonstrate that the proposed ProtoAU outperforms several representative methods.} 
% The source codes of our proposed ProtoAU are available at \url{https://github.com/oceanlvr/ProtoAU}.
\end{abstract}


\section{Introduction}
\begin{figure}[t]
    \centerline{\includegraphics[width=0.85\columnwidth]{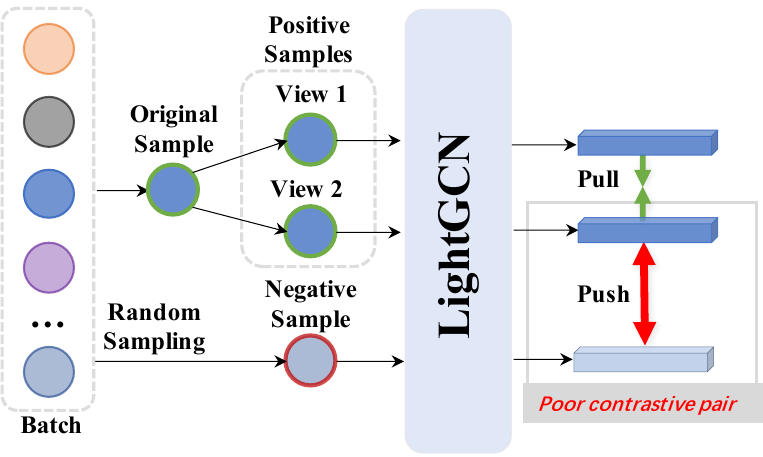}}
    \caption{Illustration of sampling bias issue. The poor contrastive pair, which contains random 
 negative samples with similar semantics to the original sample, would affect the performance of the recommendation system.}
    % \caption{Random sampling from a batch: A suboptimal sampling approach involves semantic consistency between negatives and positives, resulting in a loss of representation space.}
    \label{fig}
\end{figure}
Recommendation Systems (RS), vital for web infrastructure, play an essential role in guiding users to discover items of interest \cite{herlocker2004evaluating,ijcnn1,ijcnn2}. Among them, Collaborative Filtering (CF) recommendation stands out as one of the most popular methods for constructing personalized recommendations, particularly effective in addressing information overload scenarios. 
Graph Collaborative Filtering (GCF), the CF methods based on Graph Neural Networks (GNN), have garnered widespread attention and popularity \cite{berg2017graph,hamilton2017inductive,ijcnn3}.

GCF-based models can capture the interactions between users and items and  multi-hop neighbor relationships via graph structure, but the data sparsity inherent in recommendation systems limits the performance of these models. Additionally, the GNN aggregation in GCF may exacerbate the negative impact of data sparsity \cite{wu2021self}. Data sparsity leads to sparse supervision signals, skewed data distribution, and noises in interactions, which affect the performance of recommendation systems.
% and limitations associated with power-law distributions. User-item interactions are commonly observed in real-world scenarios, accompanied by the prevalent issues of missing or noisy data \cite{wu2021self,yao2021self}.  Moreover, user-item interactions typically exhibit \textcolor{blue}{a power-law distribution} \cite{wu2021self, yao2021self, lin2022improving}. In other words, the majority of nodes have only a few interactions, whereas a few items receive a disproportionately large number of interactions, standing in contrast to the majority that garner minimal attention.
% \textcolor{blue}{The aggregation of GNN often exacerbates the issue of data skewness in GCF.[**]} 
As a result, the Graph Contrastive Learning (GCL) approach has garnered significant attention \cite{yu2023self, wu2021self, lin2022improving}.
Specifically, many studies \cite{wu2021self, yu2021socially, cao2021bipartite, yu2021self} generate augmented views of the original graph by employing designed data augmentations, such as edge perturbation, node dropping, or sampling subgraph approach. Furthermore, these methods optimize the consistency between different augmented views by maximizing mutual information \cite{bachman2019learning}. Compared to GCF, GCL, which leverages self-supervised techniques, has the potential to be more adept at addressing data sparsity issue in recommendation systems.

% However, GCL-based methods usually suffer from the problem of sampling bias and limited semantics\cite{zhou2022debiased,robinson2020contrastive}. Specifically, the process of GCL involves random sampling from the data distribution. However, sampling negative examples with semantic similarity to the positive examples within the same batch may lead to a degradation of representations.
Despite the success of GCL-based methods, they still suffer from the problem of sampling bias and limited semantics \cite{zhou2022debiased,robinson2020contrastive}. 
As for sampling bias, the GCL process involves random sampling from the data distribution, and this approach may degrade representations when, undesirably, a semantically similar sample to the original sample is mistakenly selected as the negative sample and pushed away.
Taking Fig.~\ref{fig} as an example, the original sample and the negative sample come from the same batch, with the original sample subsequently being augmented into two positive sample views. We aim to minimize the distance between the original sample and positive sample (c.f., green arrow) while maximizing the distance between the original sample and negative sample (c.f., red arrow). 
If a sample, which is pushed away as a negative sample, is semantically close to the original sample, then the sampling bias issue occurs. 
As for limited semantics, GCL-based methods primarily focus on learning at the instance-wise level in the feature space, which might not effectively capture the global semantic information embeddings within the entire graph.  These two issues affecting the performance of GCL-based methods need to be resolved urgently.

To address the above problem, we propose the \textbf{Proto}typical contrastive learning representation from the perspective of \textbf{A}lignment and \textbf{U}niformity for recommendation (ProtoAU). Specifically, prototypical contrastive learning is proposed to ensure consistency between the prototype and its representation, it effectively modeling semantic information at the prototype level. It encourages features with similar semantics to converge towards their corresponding prototypes and, conversely, to push away from prototypes with opposing semantics. Inspired by recent work \cite{wang2022towards, wang2020understanding}, alignment and uniformity of the prototype are incorporated for optimization objectives to mitigate the issue of dimensional collapse. Specifically, we apply alignment loss to both user prototypes and item prototypes. Experiments have demonstrated that alignment and uniformity in the prototype space can effectively prevent the prototype space from falling into trivial solutions.

% The main contributions of this paper can be summarized as follows:
Key contributions of this work include:
\begin{itemize}
    \item We propose a recommendation framework that integrates prototypical contrastive learning with the objectives of alignment and uniformity, effectively addressing the data sparsity and sampling bias in recommendation systems.
    % We introduce a recommendation framework that integrates prototypical contrastive learning with alignment and uniformity objectives, thereby enhancing the accuracy of recommendation systems based on GCF models.
    % We propose a prototype contrastive learning method to capture semantic information at the prototype level, \textcolor{blue}{thus enhancing the accuracy of GCF-based recommendation systems.}
    \item We propose a prototype-level consistency loss and a corresponding loss from the perspective of alignment and uniformity, addressing sampling bias and dimensional collapse issues, respectively.
    \item We systematically analyzed the impact of various prototypes on recommendation tasks, including a detailed quantitative analysis of their effects. This provides insights into the application of prototypical contrastive learning techniques in specific recommendation tasks.
\end{itemize}

\section{Related Work}
\subsection{Graph Contrastive Learning for Recommendation}
In recent work, graph contrastive learning has been effectively integrated into GNN-based recommendation systems \cite{lin2022improving,chen2023towards,wang2022towards,wu2021self,yu2022graph,yao2021self}. The goal is to capture the supervised signal to handle the problem of data sparsity. It achieved it by discovering potential structures and maintaining consistency of information among augmented views.
For example, SGL \cite{wu2021self} generates the augmented views by randomly dropping nodes or edges. Meanwhile, it also optimizes consistency using the InfoNCE loss \cite{oord2018representation}. NCL \cite{lin2022improving} incorporates semantic prototypes and structural neighbors to improve the model's performance. LightGCL \cite{cai2023lightgcl} employs singular value decomposition to capture global-level context information for contrastive learning. SimGCL \cite{yu2022graph} proposes a feature-level perturbation augmentation method to create a contrastive view.
Although these models have achieved notable performance, the majority of them have not fully overcome the challenges associated with sampling bias and the heuristic design of hard negatives for the sampling process.

\subsection{Alignment and Uniformity}
In recent studies, many methods have indicated that the quality of representations in contrastive learning depends on two key aspects: \textit{alignment} and \textit{uniformity} \cite{wang2020understanding,zhou2022debiased}. Alignment involves bringing the features of positive pairs closer together, aiming to make them more proximate in the embedding space. It contributes to capturing similarity and relevance, enhancing the consistency among positive samples. On the other hand, uniformity, related to negative pairs, aims to distribute the features of all nodes as uniformly as possible across the unit hypersphere. In this way, the differences in features among negative samples are emphasized, helping the model better distinguish between different categories.
When employing methods that use non-negative sampling, ensuring the quality of uniformity can be challenging. One solution is to incorporate the direct optimization of uniformity and alignment in the embeddings space as objectives into the overall target function. This approach helps balance the feature learning of positive and negative samples, improving the overall model representation quality.
For example, DirectAU \cite{wang2022towards} directly optimizes the alignment and uniformity objectives attributes of user and item representations. It focuses more on improving the alignment of positive samples and ensuring a more uniform distribution of negative samples.
On the other hand, nCL \cite{chen2023towards} aims to achieve geometric features of alignment and compactness in the embedding space. This method adjusts the geometric structure of the embedding space to better capture the alignment of positive samples and ensure a more uniform distribution of negative samples, thereby enhancing the performance of contrastive learning models.

\begin{figure*}[tp]
    \centering
    \includegraphics[width=0.75\textwidth]{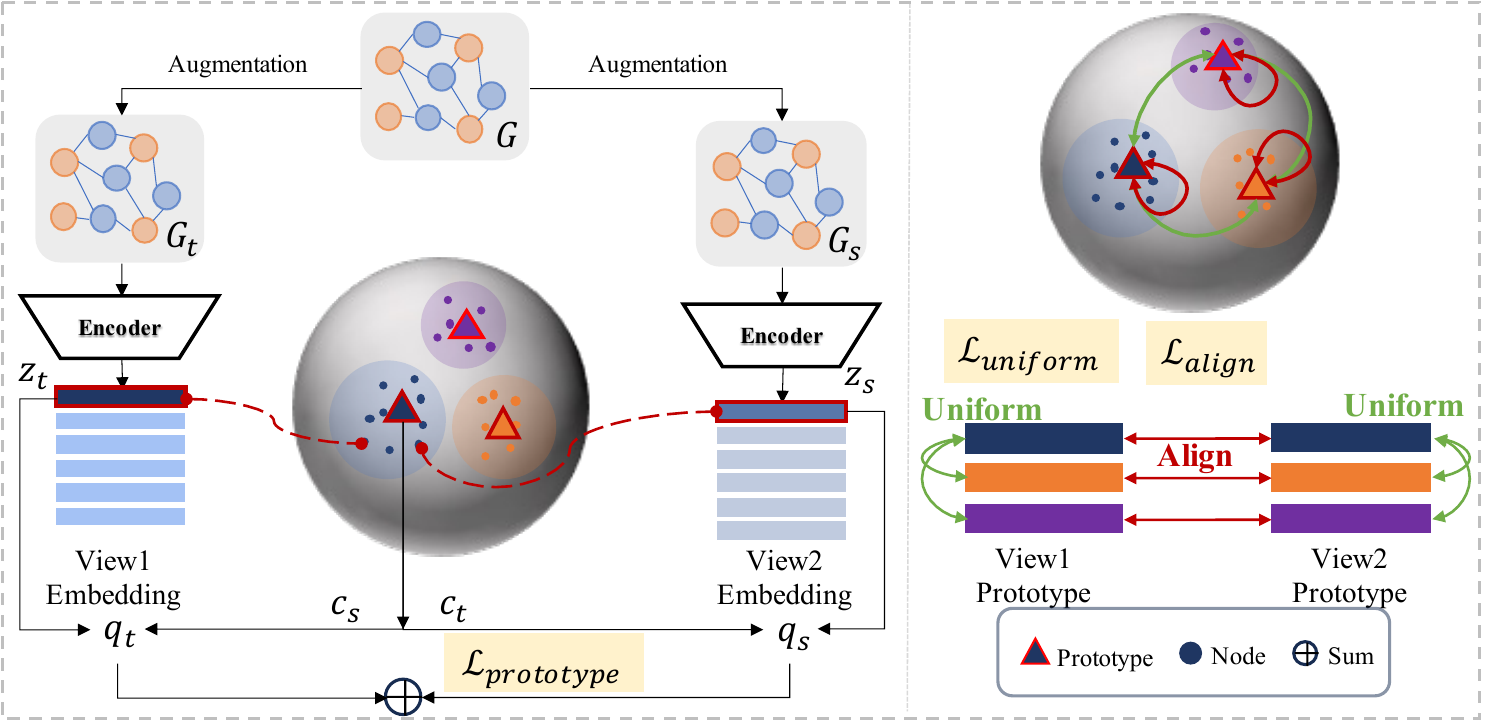}
    \caption{The overall framework of the proposed ProtoAU.}
    \label{illustrates}
\end{figure*}

\section{Methodology}

In this section, we propose a novel recommendation model called prototypical contrastive learning through alignment and uniformity (ProtoAU), illustrated in Fig. \ref{illustrates}. It is mainly composed of three parts: graph collaborative filtering backbone, prototypical contrastive learning, alignment and uniformity objectives. Specifically, ProtoAU uses a graph collaborative filtering backbone LightGCN, which generates the embeddings of user and item for the recommendation task. On top of that, ProtoAU consists of two components of the prototypical contrastive learning and the alignment and uniformity objectives for prototype training.

\subsection{Graph Collaborative Filtering Backbone}

Let $\mathcal{U}$ and $\mathcal{I}$ represent the sets of users and items, where ${u} \in \mathcal{U}$ and ${i} \in \mathcal{I}$, respectively. $R\in\{0, 1\}^{|\mathcal{U}|\times |\mathcal{I}|}$ represents a set of user-item interactions,
where $R_{u,i} = 1$ indicates that there is an interaction between user $u$ and item $i$; otherwise, $R_{u,i} = 0$.

LightGCN is used \cite{he2020lightgcn} as the base model to learn the embedding of user and item interactions. 
The base model takes the original graph as input and performs a neighborhood aggregation operation to integrate the representations of each node's adjacent nodes.
% The base model inputs the original graph $\mathcal{G}$ and aggregates the representation of its neighbors. 
The graph convolution operation is defined as shown in Eqn. \ref{lightgcn}.

\begin{equation}
\label{lightgcn}
\begin{aligned}
    e_u^{(l+1)} &= \sum_{i \in \mathcal{N} _u} \frac{1}{\sqrt{|\mathcal{N} _u| |\mathcal{N} _i|}} e_i^{(l)}, \\
    e_i^{(l+1)} &= \sum_{u \in \mathcal{N} _i} \frac{1}{\sqrt{|\mathcal{N} _i| |\mathcal{N} _u|}} e_u^{(l)},
\end{aligned}
\end{equation}    
where $e_u^{(l)}$ and $e_i^{(l)}$ denote the embeddings of user $u$ and item $i$ in the $l$-layer.
$\mathcal{N} _u$ and $\mathcal{N} _i$ represent the set of neighbors nodes of the user $u$ and item $i$, respectively.

With the embeddings of nodes in the $L$-th layer, the base model generates the representations of user and item, as shown in Eqn. \ref{euei}.

\begin{equation}\label{euei}
    e_u = \frac{1}{L+1} \sum_{l=0}^L e_u^{(l)},\qquad
    e_i = \frac{1}{L+1} \sum_{l=0}^L e_i^{(l)}.
\end{equation}

The integration of the embeddings of user $u$ and item $i$ yields a predicted preference score, i.e., the inner product of the embeddings of the user and the item, as shown in Eqn. \ref{score}.

\begin{equation}\label{score}
    \hat{y}_{u, i}=\mathbf{e}_u^{\top} \mathbf{e}_i.
\end{equation}

Meanwhile, the Bayesian Personalized Ranking (BPR) loss \cite{rendle2012bpr} is employed as the objective of the recommendation task, shown in Eqn. \ref{lossbpr}. Observed interaction data is treated as positive sample pairs, while unobserved interactions serve as negative sample pairs, aiming to maximize the gap between positive and negative samples. 
% The process of calculating the loss function is shown in Eqn. \ref{lossbpr}.

\begin{equation}\label{lossbpr}
    \mathcal{L}_{BPR}\left(\Theta\right)=-\sum_{(u, i, j) \in \mathcal{D}}\log{ \sigma\left(\hat{y}_{u, i}-\hat{y}_{u, j}\right)},
\end{equation}
where $\mathcal{D} = \{\left(u,i,j\right)|R_{u,i}=1,R_{u,j}=0\}$ denotes the training dataset. $\sigma\left(\cdot \right)$ is a sigmoid function, and $\Theta$ is the trainable mode parameter.

\subsection{Prototypical Contrastive Learning}
\label{sec:pcl}

Instance-wise graph contrastive learning can capture the invariance between two views through the InfoNCE loss \cite{oord2018representation}. Specifically, the instance-wise contrastive loss encourages alignment between representations of the same user or item and ensures uniformity in the representation space to sufficiently distinguish different users or items. Typically, the contrastive loss is expressed as illustrated in Eqn. \ref{lossinfonce}:

\begin{equation} \label{lossinfonce}
    \mathcal{L}_{InfoNCE}=-\sum_{i \in \mathcal{B}}^n \log \frac{\exp \left(\mathbf{v}_i \cdot \mathbf{v}_i^{\prime} / \tau\right)}{\sum_{j \in \mathcal{B}} \exp \left(\mathbf{v}_i \cdot \mathbf{v}_j^{\prime} / \tau\right)},
\end{equation}
where $\tau$ represents the temperature coefficient. $\mathbf{v}_i$, $\mathbf{v}_i^{\prime}$, $\mathbf{v}_j^{\prime}$ are the original, positive, and negative final representations for each sample batch $\mathcal{B}$, respectively. 

In recommendation tasks, the final representations for users and items are represented by $e_u$ and $e_i$, which correspond to $\mathbf{v}_i$ in Eqn. \ref{lossinfonce}.
% In the context of recommendation tasks, $e_u$ and $e_i$ are denoted as the final representations for users and items, replacing $\mathbf{v}_i$ accordingly. 
However, since negative samples $\mathbf{v}_j^{\prime}$ are commonly obtained through random sampling within the same batch, those methods may suffer from the sampling bias issue, as depicted in Fig.~\ref{fig}.
% Illustrated in Fig.~\ref{fig}, these methodologies are suffered from the sample bias, which causes the separation of similar nodes and, consequently, leading to a degradation in performance.
% As depicted in Fig.~\ref{fig}, the methodologies commonly employed for obtaining negative samples through random sampling within the same batch are prone to sample bias. 
The sampling bias issue leads to the separation of similar nodes, resulting in a decrease in the performance of the recommendation system.

To address the problem of sample bias, we design a prototypical contrastive learning methodology aimed at capturing the consistency between users/items and their prototypes. This involves applying clustering to the prototypes of instances, where each prototype encapsulates a group of nodes sharing similar semantic information. Specifically, considering all final user representation features $U \in \mathbb{R}^{U \times \mathcal{D}}$ and item representation features $I \in \mathbb{R}^{I \times \mathcal{D}}$, we cluster the user and item final representations into $K$ trainable prototype centroids: $C^{U} \in \mathbb{R}^{K \times \mathcal{D}}$ and $C^{I} \in \mathbb{R}^{K \times \mathcal{D}}$, respectively. $C^{U}$ comprises $K$ columns representing user prototype features, denoted as $\left\{c^{\left({1}\right)}_u,\ldots, c^{\left({K}\right)}_u \right\}$. If the user representation $e_u$ is associated with the user prototype $c^{\left(n\right)}_u$ and $e_i$ is associated with the item prototype $c_i^{\left(n\right)}$, the user prototypical contrastive loss can be described as Eqn. \ref{lossu}.

\begin{equation}\label{lossu}
    l_{proto }^U=-\sum_{n \in U} \log \frac{\exp \left(\mathbf{e}_u \cdot \mathbf{c}^{\left(n\right)}_u / \tau\right)}{\sum_{c_u^{\left(t\right)} \in C^{U}} \exp \left(\mathbf{e}_u \cdot \mathbf{c}^{\left(t\right)}_u / \tau\right)}.
\end{equation}

Similarly, the item prototype loss is described as Eqn. \ref{lossi}.

\begin{equation}\label{lossi}
    l_{proto }^I=-\sum_{n \in I} \log \frac{\exp \left(\mathbf{e}_i \cdot \mathbf{c}^{\left(n\right)}_i / \tau\right)}{\sum_{c_i^{\left(t\right)} \in C^{I}} \exp \left(\mathbf{e}_i \cdot \mathbf{c}^{\left(t\right)}_u / \tau\right)}.
\end{equation}

Thus, the overall prototypical contrastive loss can be described as Eqn. \ref{originProto}.

\begin{equation}
    \label{originProto}
    \mathcal{L}_{proto} = l_{proto }^U + l_{proto }^I.
\end{equation}

A simple prototypical contrastive learning model can be implemented by utilizing the dot product to match each instance with its closest prototype in terms of semantics.
However, recent investigations \cite{hua2021feature} have highlighted that direct optimization of the consistency between prototypes and instances often leads to an imbalance in the allocation of most instances to a few prototypes. This imbalance in allocation between prototypes and instances results in degenerate solutions.
In this paper, we formulize this issue as an optimal transport problem, aiming to maximize the grouping of samples that exhibit higher inner products with their corresponding prototypes. Specifically,
the problem is defined as assigning samples within a batch to $K$ clusters according to a distribution probability $q$, while ensuring constraints regarding inter-class equilibrium.
% we propose a method involving the distribution of samples within a batch across $K$ clusters based on a distribution probability $q$, while ensuring constraints regarding inter-class equilibrium. 
To illustrate, considering the user as an example, we have a user feature vector $Z_u=[e_u^{\left(1\right)},\ldots,e_u^{\left(U\right)}]$ and their corresponding prototype vector $C_u=[c_u^{\left(1\right)},\ldots,c_u^{\left(K\right)}]$. The computation of the optimal distribution probability is denoted as $Q_u=[q_u^{\left(1\right)},\ldots,q_u^{\left(U\right)}]$.

% This scenario is approached as an optimal transport problem, aiming to maximize the grouping of samples that exhibit higher inner products with their corresponding prototypes.

% We implemented the basic paradigm of prototypical contrastive learning by applying dot product to match its closest prototype in semantics.
% However, recent studies indicate \cite{hua2021feature} that directly optimizing the consistency between prototypes and instances leads to the imbalance allocation of most instances to a few prototypes.
% The imbalance in the allocation between prototypes and instances results in degeneracy solution.
% This involves distributing samples within a batch across $K$ clusters in accordance with distribution probability $q$,
% while adhering to constraints pertaining to inter-class equilibrium.
% Intuitively, we take user as the example, given $U$ feature vector $Z_u=[e_u^{\left(1\right)},\ldots,e_u^{\left(U\right)}]$ and their corresponding prototypes vector $C_u=[c_u^{\left(1\right)},\ldots,c_u^{\left(K\right)}]$.
% We denote the computation of the optimal distribution probability as $Q_u=[q_u^{\left(1\right)},\ldots,q_u^{\left(U\right)}]$.
% The scenario is approached as an \textit{optimal transport problem}.
% The objective is to maximize the grouping of samples that exhibit a higher inner product with their corresponding prototypes.

\begin{equation}
\begin{aligned}
    \max _{\mathbf{Q} \in \mathcal{Q}_u} \operatorname{Tr}\left(\mathbf{Q}_u^{\top} \mathbf{C}_u^{\top} \mathbf{Z}_u\right)+\varepsilon H(\mathbf{Q}_u),\\
    \mathcal{Q}_u=\left\{\mathbf{Q}_u \in \mathbb{R}_{+}^{K \times B} \left\lvert\, \mathbf{Q_u} \mathbf{1_U}=\frac{1}{K} \mathbf{1_K}\right., \mathbf{Q}^{\top} \mathbf{1}_K=\frac{1}{U} \mathbf{1}_U\right\},
\end{aligned}
\end{equation}
where $\mathbf{1_K}$ represents a vector of ones in $K$ dimensions. $H(Q) = -\sum_{ij} Q_{ij} \log Q_{ij}$ denotes the entropy function, which governs the dispersion of distribution probabilities.

Then, the approximate probability $Q$ is determined using the Sinkhorn-Knopp algorithm \cite{cuturi2013sinkhorn}.

\begin{equation}
\label{eqn:sinkhorn}
Q_u = \text{Diag}(\mu ) \exp\left( \frac{C_u^\top Z_u}{\varepsilon} \right) \text{Diag}(\nu ),
\end{equation}
where $\mu \in \mathbb{R}^K$ and $\nu \in \mathbb{R}^U$ are renormalization vectors.

For augmented views $G_t$ and $G_s$ of the original graph $G$, we can compute their respective distribution transport codes $q_t$ and $q_s$, as they represent different views $z_t$ and $z_s$ of the same user, as illustrated in Fig.~\ref{illustrates}. Subsequently, we exchange the distribution transport prediction tasks according to Eqn. \ref{originProto}, utilizing the cross-entropy loss depicted in Eqn. \ref{align}.

\begin{equation}
\begin{aligned}
\label{swap}
    \mathcal{L}\left(z_t,q_s\right)=-\sum_K q_s^{(k)} \log p_t^{(k)},\\
    p_t^{(k)} =  \frac{\exp(z_t \cdot c_k / \tau)}{\sum_{c^{k^{\prime}}} \exp(z_t \cdot c^{k^{\prime}} / \tau)}.
\end{aligned}
\end{equation}

Correspondingly, the following user prototypical contrastive learning loss (as shown in Eqn. \ref{lossprotou}) is derived through the integration of the symmetrical component.

\begin{equation}
\begin{aligned} \label{lossprotou}
    \mathcal{L}_ {proto}^U = \mathcal{L}\left(z_t,q_s\right) + \mathcal{L}\left(z_s,q_t\right).
\end{aligned}
\end{equation}

Similarly, the overall prototypical contrastive learning loss for items can be derived as shown in Eqn. \ref{lossproto}.
\begin{equation}\label{lossproto}
\begin{aligned}  
    \mathcal{L}_{proto} = \mathcal{L}_{proto}^U+\mathcal{L}_{proto}^I.
\end{aligned}
\end{equation}

\subsection{Alignment and Uniformity Objectives}
The absence of explicit negative examples suggests that directly optimizing the consistency loss between an instance and its prototype may result in potential dimensional collapse issues at the instance-prototype level.
Given that Eqn. \ref{swap} introduces cross-prototype contrast to our task, we should acknowledge the possibility of dimensional collapse issues occurring in the prototype space.
Recent studies \cite{wang2020understanding, zhou2022debiased} have demonstrated that \textit{alignment} and \textit{uniformity} are two effective principles within contrastive learning. Many works have directly optimized these metrics, leading to impressive results in recommendation systems \cite{chen2023towards, wang2022towards}. Motivated by this, we propose alignment and uniformity objectives for prototypical contrastive learning to further mitigate the risk of dimensional collapse issues.

\textbf{Alignment} minimizes the distance between an original 
sample and a positive sample. In practice, it ensures that two samples forming a positive pair preserve invariant information. 
The alignment, as implemented in the prototype representation space, effectively reduces the distances between prototype representations of similar instances, thereby enhancing the coherence of the embedding space, as detailed in Equation \ref{lossalign}.
% The alignment is employed in the prototype representation space to ensure that prototype representations of similar instances are brought closer together in the embedding space, which is illustrated in Eqn. \ref{lossalign}.

\begin{equation}\label{lossalign}
{l}_{align}(c_t,c_s; \alpha) = \mathbb{E}_{(t,s) \sim P_{\text{positive}}} \left[ \| c_t - c_s \|_2^{\alpha} \right],
\end{equation}
where $c_t$ and $c_s$ are the prototype centroids of the positive sample pairs $t$ and $s$, respectively. $\alpha$ is the distance factor between positive examples.

The consistency loss for users and item prototypes can be represented as Eqn. \ref{align}.

\begin{equation}
    \label{align}
    \mathcal{L}_{align} = {l}_{align}^U + {l}_{align}^I.
\end{equation}

\textbf{Uniformity} ensures that sample representations are well-separated on the hypersphere in contrastive learning. It implements a repulsive force between non-similar instance representations through the utilization of a Gaussian potential kernel, effectively dispersing them evenly across the surface of the hypersphere. This mechanism ensures that representations do not cluster too closely together, mitigating the dimensional collapse issue and promoting a diverse and generalized feature space that aids in better discrimination of novel instances. Inspired by this foundation, we propose its application to the prototype space, ensuring spatial uniformity under inter-prototype conditions.

\begin{equation}
    l_{uniform}(c; \beta) = \mathbb{E}_{(t,s) \sim P_{proto}} \left[ e^{-\beta\| c_t - c_s \|_2^2} \right],
\end{equation}
where $c_t$ and $c_s$ are the prototype centroids that sample $i$ belongs to. $\beta$ is a scaling factor that regulates the degree of repulsion between different prototypes.

Under recommendation system tasks, uniformity can be expressed as Eqn. \ref{uniform}.

\begin{equation}
\label{uniform}
    \mathcal{L}_{uniform} = {l}_{uniform}^U + {l}_{uniform}^I.
\end{equation}

\subsection{Overall Loss}
The overall loss of the proposed ProtoAU consists of all the aforementioned losses, as illustrated in Fig.~\ref{illustrates}. We employ a multi-task learning strategy to simultaneously train those loss, as referenced in \ref{lossbpr}.

\begin{equation}
    \label{totalLoss}
    \mathcal{L}_{ProtoAU} =\mathcal{L}_{BPR} + \lambda_1 \mathcal{L}_{proto} + \lambda_2 \mathcal{L}_{align} + \lambda_3 \mathcal{L}_{uniform},
\end{equation}
where $\lambda$ are the joint task weights.

\begin{algorithm}
    \label{psudo}
    \caption{ProtoAU in Pytorch-like style.}
    \SetAlgoLined
    \DontPrintSemicolon
    \SetKwBlock{With}{with torch.no\_grad():}{} % Remove 'end' from 'With' block
    
    \SetKwProg{Fn}{Fn}{:}{}
    \SetKwFunction{FAlign}{Alignment}
    \SetKwFunction{FUniform}{Uniformity}
    \SetKwFunction{FMain}{Sinkhorn}
    
    \KwData{C: prototypes; \(x_t, x_s\): augmented view}
    
    \For{\(x_t, x_s\) in loader}{
        \( z = \text{model(concat}(x_t, x_s)) \) \;
        scores = mm(z, C) \;
        \(\text{scores}_t\) = scores[:N]
        \(\text{scores}_s\) = scores[N:] \;
        \(C_t\) = C[\(\text{scores}_t\).max(dim=1)[1]]\;
        \(C_s\) = C[\(\text{scores}_s\).max(dim=1)[1]]\;
        \With{ \tcp{Optimal transport code}
            \(q_t = \text{Sinkhorn}(\text{scores}_t) \) \tcp{Eqn. \ref{eqn:sinkhorn}} 
            \(q_s = \text{Sinkhorn}(\text{scores}_s) \) \;
        }
        \( p_t = \text{Softmax}(\text{scores}_t / \text{temp}) \) \;
        \( p_s = \text{Softmax}(\text{scores}_s / \text{temp}) \) \;
        % \tcp{Prototype contrastive learning loss}
        \( L_\text{proto} = -\frac{1}{2} \times \text{mean}(q_s \times \log(p_t) + q_t \times \log(p_s)) \) \;
        \tcp{ $\mathcal{L}_{\text{bpr}}$ Computed by Eqn.\ref{lossbpr} }
        $\mathcal{L}_{\text{align}} = $ \FAlign{$C_t$, $C_s$} \tcp{Eqn.\ref{align}}
        $\mathcal{L}_{\text{uniform}} = $ \FUniform{$C$} \tcp{Eqn.\ref{uniform}}
        $\mathcal{L}_\text{ProtoAU} = \mathcal{L}_\text{BPR} + \alpha * \mathcal{L}_\text{proto} + \beta * \mathcal{L}_\text{align} + \gamma * \mathcal{L}_\text{uniform}$\;
        update(params) \;
        \With{
            \( C = \text{normalize}(C, \text{dim}=0, p=2) \)
        }
    }
\end{algorithm}

\section{Experiment}

\begin{table}[htbp]
    \caption{Datasets Statistics}
    \centering
    \begin{tabular}{lccccc}
    \toprule
    \textbf{Datasets} & \textbf{\#Users} $|\mathcal{U}|$ & \textbf{\#Items} $|\mathcal{I}|$& \textbf{\#Interactions}  $|\mathcal{R}|$ & \textbf{Density }\\ \midrule
    \textbf{Amazon-Books} & 52,463  & 91,599  & 2,984,108 &  0.00075 \\
    \textbf{iFashion}     & 300,000 & 81,614  & 1,607,813 &  0.00007 \\
    \textbf{MovieLens-1M} & 6,040   & 3,629   & 836,478   &  0.03816 \\
    \textbf{Yelp2018}     & 31,668  & 38,048  & 1,561,406 &  0.00130 \\
    \bottomrule
    \end{tabular}
    \label{table:dataset}
\end{table}

\begin{table*}[ht]
\caption{Overall Performance Comparison of Different Models}
\label{table:perf}
\begin{center}
\begin{tabular}{lccccccccc}
\toprule
Dataset & Metric & NGCF  & BPRMF & LightGCN & SGL & NCL & ProtoAU & Improvement. \\
\midrule
Amazon-Books & Recall@10  & 0.0617 & 0.0607 & 0.0801 & 0.0898 & 0.0933 & \textbf{0.0955} & +2.40\% \\
& NDCG@10 & 0.0427  & 0.0430 & 0.0565 & 0.0645 & 0.0679 & \textbf{0.0681} & +3.02\% \\
& Recall@20 & 0.0978 & 0.0956 & 0.1206 & 0.1331 & 0.1381 & \textbf{0.1383} & +1.53\% \\
& NDCG@20 & 0.0537  & 0.0537 & 0.0689 & 0.0777 & 0.0815  & \textbf{0.0818} & +2.25\% \\
& Recall@50 & 0.1699 & 0.1681 & 0.2012 & 0.2157 & 0.2175 & \textbf{0.2202} & +1.24\% \\
& NDCG@50 & 0.0725 & 0.0726 & 0.0899 & 0.0992 & 0.1024 & \textbf{0.1040} & +1.56\% \\
\midrule
iFashion & Recall@10 & 0.0382   & 0.0303 & 0.0457 & 0.0461 & 0.0477 & \textbf{0.0488} & +2.31\% \\
& NDCG@10 & 0.0198  & 0.0161 & 0.0246 & 0.0248 & 0.0259 & \textbf{0.0266} & +2.70\% \\
& Recall@20 & 0.0615 & 0.0467 & 0.0692 & 0.0692 & 0.0713 & \textbf{0.0723} & +1.40\% \\
& NDCG@20 & 0.0257  & 0.0203 & 0.0246 & 0.0307 & 0.0319 & \textbf{0.0325} & +1.89\% \\
& Recall@50 & 0.1081   & 0.0799 & 0.1144 & 0.1141 & 0.1165 & \textbf{0.1181} & +1.37\% \\
& NDCG@50 & 0.0349   & 0.0269 & 0.0396 & 0.0396 & 0.0409 & \textbf{0.0421} & +2.93\% \\
\midrule
MovieLens-1M
& Recall@10 & 0.1846  & 0.1804 & 0.1876 & 0.1888 & 0.2057 & \textbf{0.2185} & +6.22\% \\
& NDCG@10 & 0.2528   & 0.2463 & 0.2514 & 0.2526 & 0.2732 & \textbf{0.2888} & +5.71\% \\
& Recall@20 & 0.2741  & 0.2714 & 0.2796 & 0.2848 & 0.3037 & \textbf{0.3192} & +5.13\% \\
& NDCG@20 & 0.2614   & 0.2569 & 0.2620 & 0.2649 & 0.2843 & \textbf{0.3021} & +6.26\% \\
& Recall@50 & 0.4341   & 0.4300 & 0.4469 & 0.4487 & 0.4686 & \textbf{0.4848} & +3.45\% \\
& NDCG@50 & 0.3055   & 0.3014 & 0.3091 & 0.3111 & 0.3300 & \textbf{0.3439} & +4.21\% \\
\midrule
Yelp2018
& Recall@10 & 0.0630  & 0.0643 & 0.0730 & 0.0833 & 0.0920 & \textbf{0.0975} & +5.98\% \\
& NDCG@10 & 0.0446   & 0.0458 & 0.0520 & 0.0601 & 0.0678 & \textbf{0.0718} & +5.90\% \\
& Recall@20 & 0.1026  & 0.1043 & 0.1163 & 0.1288 & 0.1377 & \textbf{0.1407} & +2.18\% \\
& NDCG@20 & 0.0567   & 0.0580 & 0.0652 & 0.0739 & 0.0817 & \textbf{0.0860} & +5.30\% \\
& Recall@50 & 0.1862  & 0.1862 & 0.2016 & 0.2140 & 0.2247 & \textbf{0.2285} & +1.69\% \\
& NDCG@50 & 0.0784   & 0.0793 & 0.0875 & 0.0964 & 0.1046 & \textbf{0.1066} & +1.85\% \\
\bottomrule
\end{tabular}
\end{center}
\end{table*}

In this section, we first introduce the dataset. Then, we conduct a series of experiments to verify the performance of the proposed ProtoAU on the task of recommendation and visualize the representation quality. Finally, we test the sensitivity of the parameters.

\subsection{Experiments Settings}

\subsubsection{Datasets}
To assess the effectiveness of the proposed ProtoAU model, we conduct several experiments on four publicly available datasets: Yelp2018, Amazon-Books \cite{mcauley2015image}, iFashion\cite{chen2019pog}, and MovieLens-1M\cite{harper2015movielens}.
Consistent with standard practices in prior work,
we exclude users and items with fewer than 15 interactions from the Yelp2018 and Amazon-Books datasets and eliminate duplicate interactions from the original data.
The detailed core configurations of the datasets are presented in Table~\ref{table:dataset}.
For public dataset, we randomly shuffle and split it into training, testing, and validation subsets with the ratio of 8:1:1.

\subsubsection{Baselines}

In our comparison with established methods, our proposed model ProtoAU is benchmarked against the following baseline models:

\begin{itemize} 
    \item \textbf{NGCF} \cite{wang2019neural} utilizes graph neural networks to enhance user-item interaction embeddings, capturing collaborative filtering effects effectively.
    % \item \textbf{BUIR} \cite{lee2021bootstrapping} disentangles user-item interaction graphs into multiple latent factors for more interpretable and effective representation learning in recommendations.
    \item \textbf{BPRMF} \cite{rendle2012bpr} based on matrix factorization, applies a pairwise ranking loss to personalizing the recommendation process.
    \item \textbf{LightGCN} \cite{he2020lightgcn} simplifies Graph Convolutional Networks by focusing on linear propagation of collaborative signals, enhancing both efficiency and effectiveness.
    \item \textbf{SGL} \cite{wu2021self} incorporates self-supervised learning into recommendation systems, and it notably using SGL-ED for enhanced model performance.
    \item \textbf{NCL} \cite{lin2022improving} introduces a novel architecture for collaborative learning, aiming to capture complex user-item interactions for improved recommendation accuracy.
\end{itemize}

\subsubsection{Evaluation Metrics}

In measuring the performance of our recommendation system, we use metrics: $\text{Recall}@K$ and $\text{NDCG}@K$ where $K$ includes $[10, 20, 50]$. 

\subsubsection{Hyper-parameter}
To ensure a fair comparison, we optimal hyperparameter settings as reported in the baseline models' original papers and subsequently apply grid search to fine-tune all baseline hyperparameters.
Besides, 
% we use the Xavier initialization for embedding layers.
We chosen embedding size is $64$ and the batch size is $4096$.
All models are optimized by Adam optimizer with a learning rate of $0.001$.

\subsection{Overall Performance}

Table~\ref{table:perf} shows the performance of our proposed ProtoAU model compared with other baseline methods across four datasets.
Following are the insights derived from the analysis:

\begin{itemize} 
    \item Compared to traditional GNN-based collaborative filtering systems like NGCF, BPRMF, and LightGCN, the contrastive learning approach like SGL, NCL, ProtoAU achieve superior results.
    \item Compared to SGL, which pursues consistency at the graph level through node dropping / edge perturbations, NCL capture context information through neighbor nodes, achieving better representational performance across various datasets.
    % ProtoAU在各个数据集均优于其他baseline，相比于NCL去关联结构节点和语意邻居节点的一致性，我们直接去寻找语意上可能存在潜在联系的聚类，并且可以看出对比直接设计节点级别的语意领居来指导表示，alignment和uniformity组件能够更加简单有效地优化我们的原型空间
    \item ProtoAU outperforms other baselines in those datasets. Compared to its backbone LightGCN, ProtoAU achieved a $14.16\%$ improvement in recall@20 on the MovieLens-1M dataset. ProtoAU directly identifies clusters that may have latent semantic connections. Furthermore, the components of alignment and uniformity objectives can more simply and effectively optimize our prototype embedding space. Notably, on the MovieLens-1M dataset, it achieved by $31.92\%$ and $30.21\%$ in the Recall@20 and NGCG@20 metrics, respectively, which is an improvement of $5.13\%$ and $6.22\%$ over the best baseline NCL. 
\end{itemize}

% \subsection{Study of ProtoAU}
% \subsection{Impact of Data Sparsity}
\subsection{Ablation Study}

\begin{table}[ht]
\caption{The ablation study of ProtoAU and its variant.}
\begin{center}
\begin{tabular}{lcc}
\toprule
Method & \multicolumn{1}{c}{Yelp2018} & \multicolumn{1}{c}{MovieLens-1M} \\ 
& \multicolumn{1}{c}{Recall@20} & \multicolumn{1}{c}{Recall@20}\\ 
\midrule
LightGCN & 0.1163 & 0.2796 \\
\midrule
\textit{ProtoAU-w/o-Proto}  & 0.1298 & 0.3014 \\
\textbf{\%Improv} & \textbf{+11.60\%} & \textbf{+7.79\%} \\
\midrule
\textit{ProtoAU-w/o-AU}  & 0.1362 & 0.3027 \\
\textbf{\%Improv} & \textbf{+17.11\%} & \textbf{+8.26\%} \\
\midrule
ProtoAU & 0.1407 & 0.3192 \\
\textbf{\%Improv} & \textbf{+20.98\%} & \textbf{+14.16\%} \\
\bottomrule
\end{tabular}
\end{center}
\label{table:ablation}
\end{table}

In this section, we conduct an ablation study of the ProtoAU model to assess the impact of its components.
To evaluate the effectiveness of the two components, we create two variations of ProtoAU: one without prototypical contrastive learning (referred to as \textit{ProtoAU-w/o-Proto}), and the other without alignment and uniformity objectives (referred to as \textit{ProtoAU-w/o-AU}).
Specifically, in the \textit{ProtoAU-w/o-Proto}  version, the prototypical contrastive loss \ref{lossproto} is removed, and the original InfoNCE loss \ref{lossinfonce} is employed instead.
Meanwhile, in the \textit{ProtoAU-w/o-AU} version, the losses associated with the alignment \ref{align} and uniformity \ref{uniform} loss of prototypes  are removed.
The result are shown in Table.~\ref{table:ablation}.
Analyzing the results, it is observed that removing prototypical contrastive learning (\textit{ProtoAU-w/o-Proto}) leads to a performance decrease compare to original model ProtoAU, yet it still outperforms its backbone, LightGCN.
Conversely, the removal of alignment and uniformity objectives (\textit{ProtoAU-w/o-AU}) indicates that these attributes effectively explore better representations within the prototype representation, further enhancing performance on top of the prototypical contrastive component.

% \subsubsection{Effectiveness of Prototypical Contrastive Learning}
% 仅仅使用INFONCE优化原型和实例
% \subsubsection{Effectiveness of Alignment and Uniformity}
% 把Alignment and Uniformity损失丢了看效果

% 按交互多少分布 分组实验
\subsection{Visualizing Representations}

\begin{figure}[t]
\centering
\subfigure[\textbf{LightGCN}]{
    \label{fig:visualize:lightgcn}
    \includegraphics[width=0.28\linewidth]{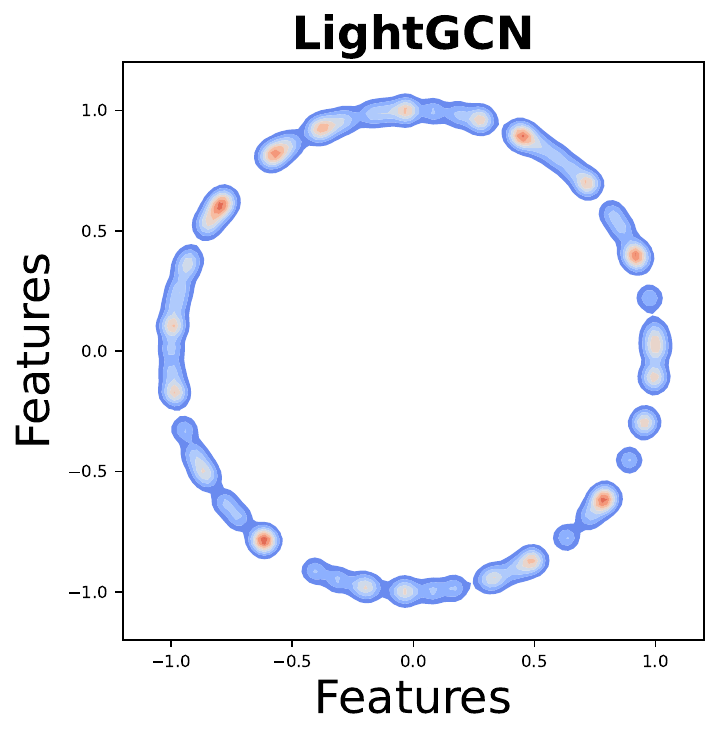}
}
\subfigure[\textbf{NCL}]{
    \label{fig:visualize:ncl}
    \includegraphics[width=0.28\linewidth]{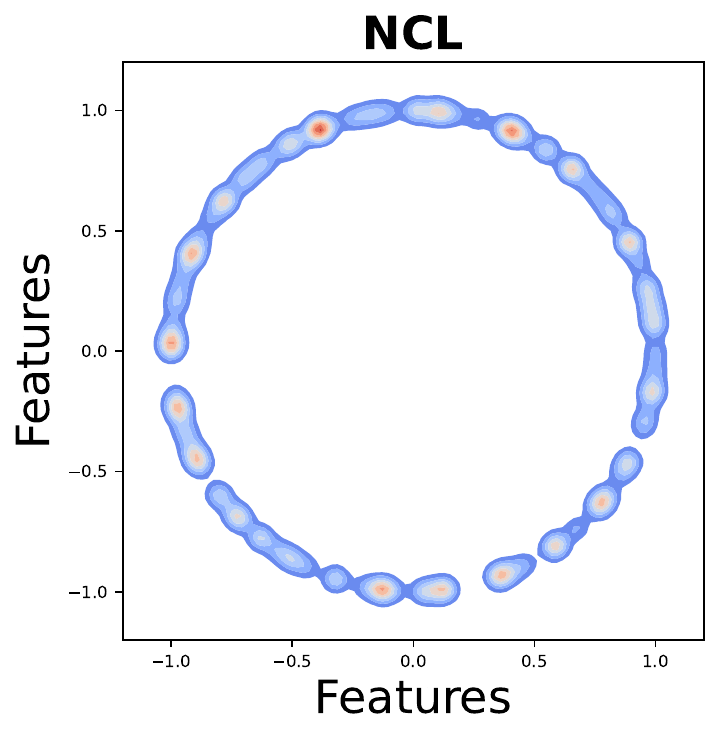}
}
\subfigure[\textbf{ProtoAU}]{
    \label{fig:visualize:protoau}
    \includegraphics[width=0.28\linewidth]{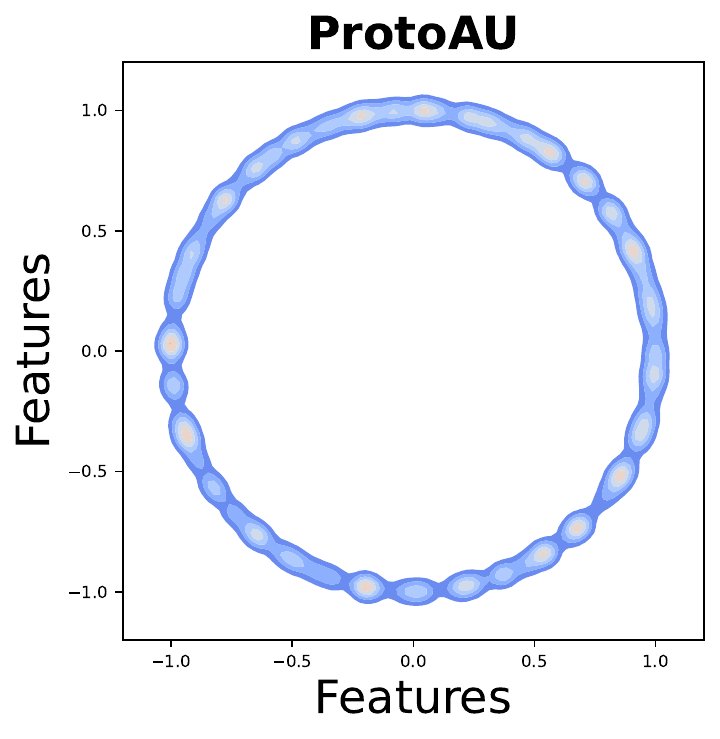}
}
\caption{Visualizing study for the embeddings in recommendation systems on datasets Yelp2018.}
\label{fig:visualize}
\end{figure}

This study evaluates the quality of item embeddings generated by LightGCN, NCL, and our proposed ProtoAU in recommendation systems, focusing on the uniformity of embeddings in the spatial domain to elucidate their performance. The experiment's objective is to visualize and analyze the quality of item embeddings from each method  on the unit hypersphere by using t-SNE\cite{van2008visualizing} on datatset Yelp2018, hypothesizing that uniformity is a critical factor in their effectiveness. Utilizing the consistency between embedding uniformity and contrastive learning objective~\cite{lin2022improving}, we visualized the spatial distributions of embeddings for all three methods, as shown in Fig. \ref{fig:visualize}. The results indicate that ProtoAU's embeddings are more uniformly distributed than those from NCL and LightGCN, suggesting a more effective representation of diverse user preferences and item characteristics. This superiority of ProtoAU~\ref{fig:visualize:protoau} is attributed to its incorporation of prototypical contrastive learning elements, alignment, and uniformity objectives, enhancing its modeling capacity and addressing dimensional collapse more effectively than NCL~\ref{fig:visualize:ncl}. Consequently, this study demonstrates ProtoAU's potential for superior performance in recommendation systems through more uniformly distributed item embeddings.

% t-nse 消融
\subsection{Hyper-parameter Analysis}

% 参数检验
\subsubsection{Impact of $K$}

In this experiment, we investigate the impact of the number of prototypes $K$ on the performance of recommendation systems MovieLens-1M dataset.
As shown in Fig. \ref{fig:sen-1}, we compare the performance of ProtoAU, which incorporates prototypical contrastive learning, with its backbone model, LightGCN, to analyze the effect of this learning approach.
The experimental results on MovieLens-1M dataset indicate that ProtoAU consistently outperforms LightGCN, with optimal performance achieved when $K$ is around 1500.
Furthermore, when $K$ is greater than or equal to 0, a significant performance improvement is observed for ProtoAU, which can be attributed to the diverse aggregation of numerous prototypes, bringing about significant benefits.

\subsubsection{Impact of $\lambda_2$ and $\lambda_3$}
\begin{figure}[t]
\centering
\subfigure[The number of prototypes.]{
    \label{fig:sen-1}
    \includegraphics[width=0.45\linewidth]{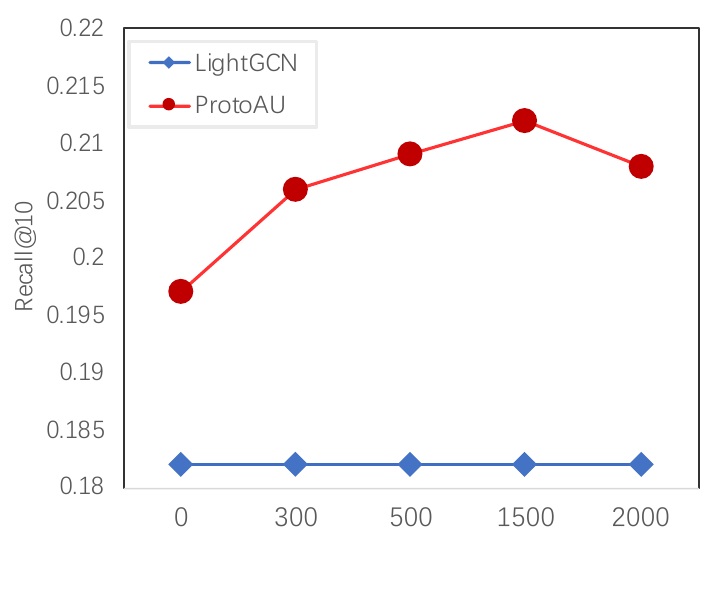}
}
\subfigure[The weights of alignment uniformity objectives.]{
    \label{fig:sen-2}
    \includegraphics[width=0.45\linewidth]{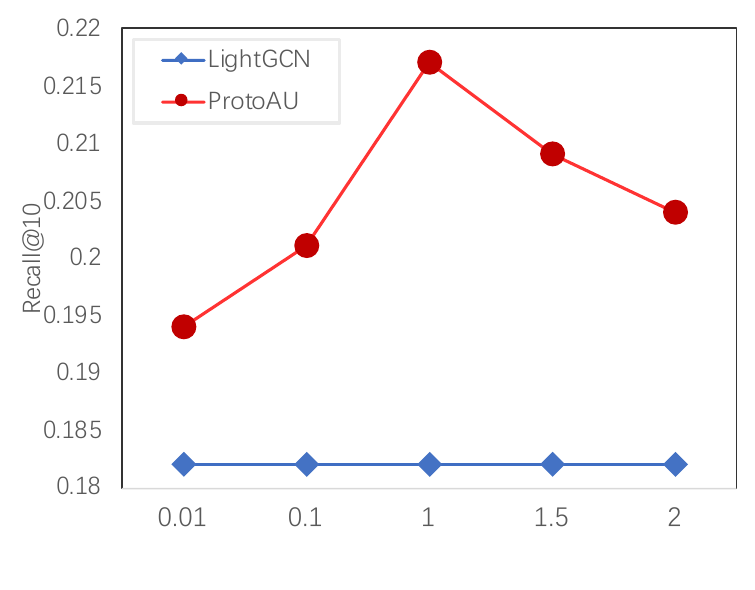}
}
\caption{Sensitivity analysis of the number of prototypes $K$ and the weights of alignment uniformity objectives $\lambda_2$, $\lambda_3$ on MovieLens-1M dataset with Recall\@10 results presentation.}
\label{fig:convergence}
\end{figure}

This experiment investigates the impact of the weight of the alignment objective $\lambda_2$,
and the weight of the uniformity objective $\lambda_3$, on prototypical contrastive learning.
We compared the performance of ProtoAU with varying $\lambda_2$ and $\lambda_3$ on MovieLens-1M dataset.
The results, as shown in Fig. \ref{fig:sen-2},
illustrate the relative importance of alignment and uniformity via the $\frac{\lambda_3}{\lambda_2}$ metric on the x-axis, e.g., $\frac{\lambda_3}{\lambda_2} > 1$ suggests that prototype training places more emphasis on uniformity over alignment. The results indicate that ProtoAU with $\frac{\lambda_3}{\lambda_2} = 1$ achieves optimal performance $21.72\%$ on Recall@10, demonstrating that the recommendation task on the MovieLens-1M dataset requires a balance between alignment and uniformity. Future work can also utilize this approach when setting the weights of the alignment and uniformity objectives.

\section{Conclusion}
% In this paper, we propose ProtoAU, a representation learning method for recommendation. The key focus of ProtoAU lies in its ability to capture intricate relationships between user and item interactions. Moreover, ProtoAU ensures consistency across different augmentations from the origin graph, aiming to eliminate the need for random sampling of contrastive pairs. It also enables the prevention of dimensional collapse problems by introducing alignment and uniformity. Experimental results demonstrate the importance of incorporating the prototypical contrastive learning, alignment and uniformity into the model.
In this paper, we propose ProtoAU, a representation learning method for recommendation that excels in capturing intricate relationships between user and item interactions. ProtoAU ensures consistency across different augmentations from the origin graph, eliminating the need for random sampling of contrastive pairs, and introduces alignment and uniformity objectives to prevent dimensional collapse problems. Experimental results underscore the significance of incorporating prototypical contrastive learning, alignment, and uniformity into the model. Building on these findings, future work could extend the application of ProtoAU to more specific recommendation tasks, further exploring the potential of prototypes in enhancing recommendation accuracy and user satisfaction.

\newpage

\bibliography{reference}
\bibliographystyle{IEEEtran}

\end{document}